# Context Tree for Adaptive Session-based Recommendation

Fei Mi
Artificial Intelligence Lab

Boi Faltings
Artificial Intelligence Lab

École polytechnique fédérale de Lausanne, Switzerland
{firstname.lastname}@epfl.ch

École polytechnique fédérale de Lausanne, Switzerland
{firstname.lastname}@epfl.ch

## ABSTRACT

There has been growing interests in recent years from both practical and research perspectives for session-based recommendation tasks as long-term user profiles do not often exist in many real-life recommendation applications. In this case, recommendations for user's immediate next actions need to be generated based on patterns in anonymous short sessions. An often overlooked aspect is that new items with limited observations arrive continuously in many domains (e.g. news and discussion forums). Therefore, recommendations need to be adaptive to such frequent changes. In this paper, we benchmark a new nonparametric method called context tree (CT) against various state-of-the-art methods on extensive datasets for session-based recommendation task. Apart from the standard static evaluation protocol adopted by previous literatures, we include an adaptive configuration to mimic the situation when new items with limited observations arrives continuously. Our results show that CT outperforms two best-performing approaches (recurrent neural network; heuristic-based nearest neighbor) in majority of the tested configurations and datasets. We analyze reasons for this and demonstrate that it is because of the better adaptation to changes in the domain, as well as the remarkable capability to learn static sequential patterns. Moreover, our running time analysis illustrates the efficiency of using CT as other nonparametric methods.

## 1 INTRODUCTION

With the increased availability of data, machine learning and deep learning have become the method of choice for knowledge acquisition in intelligent systems and various applications. However, new data often arrives continuously, therefore, the knowledge derived from it often looses its value over time. Therefore, machine learning models should constantly acquire new knowledge and adapt to dynamic environments. This situation, referred to adaptive environment in our paper, is very common in recommendation domains. Two aspects require practical recommendation system to be used in adaptive environment. First, user interests and preferences drift fast as more items are generated [18]. This is especially true for social networks or news websites where recommendations need be adaptive to drifting trends rather than recommending obsolete or well-known information. Second, the pool of items to be recommended and items themselves are evolving over time because items can be added and updated very frequently in practical recommendation applications like discussion forums. Therefore, updating models efficiently and adapting to new items are crucial. Traditional recommendation techniques, such as collaborative filtering and methods based on matrix factorization, build an increasingly complex model of users and items. Therefore, when a new item is superseded by a newer version or a new preference pattern appears, it takes time and data for recommendations to adapt.

In recent years, session-based recommendation starts to get more practical attention since many small retailers and media sites do not track the identity of users over a long period of time. In this case, recommendations are generated using only browser fingerprints, and the typical task is to predict the next item a user is going to consume from a session of actions from the past. Two state-of-the-art methods so far for session-based recommendation task make use of either sequential patterns [12, 13] or co-occurrence statistics [16] from sessions of activities to build recommendation models. However, they are all evaluated in a static configuration where the model is trained on a large training set and evaluated on a small testing set with only items from the training set. While such a static configuration is standard for testing offline machine learning models, we contend that it is crucial to test the capabilities of recommendation algorithms in an adaptive configuration where new items and patterns arrive continuously with limited observations. The challenge is combining old knowledge with new patterns such that both can be identified fast and accurately in adaptive environments. This is especially important for session-based recommendation since only relatively short-term session data is used to train a model. Therefore, we conduct experiments in both static and adaptive configurations to test the capability and usability of recommendation models.

Session-based recommendation can be formulated as a standard sequence learning problem and the state-of-the-art method for sequence learning is the recurrent neural network (RNN) [22], which is broadly devised for language modeling [28], sequence prediction [8], as well as session-based recommendation [12, 13]. Special recurrent units like Long Short-term Memory (LSTM) [14] or Gated Recurrent Units (GRU) [4] are widely applied to RNNs for modeling long-term dependencies within a sequence with the help of specialized gate functions, which makes RNNs good at learning sequences with sufficient training data and fixed set of labels. However, new items arrive frequently with limited observations in adaptive environments. This makes updating an RNN difficult because of the large number of parameters that parametrize an RNN on new items.

Nonparametric methods [9] are ideal candidates for dynamic environments since the amount of information a nonparametric model captures evolve as the amount of data grows. Moreover, the efficiency of computing and updating makes nonparametric methods easy to use in industry. For session-based recommendation task, [16] proposes a strong nonparametric method very recently using session-based nearest-neighbor methods and they show that it outperforms state-of-the-art models based on RNNs with better computation efficiency. In this paper, we propose another category of nonparametric method, based on variable-order Markov models. More specifically, we use a structure called context tree (CT) [32] which was originally proposed for lossless data compression. [1, 5, 19] applied this structure to various discrete sequence prediction tasks. Recently it was applied to news recommendation by [6, 7]. Through extensive comparisons with other methods, we show that this structure elegantly combines old and new knowledge so as to be an effective and efficient choice for both static and adaptive configurations for session-based recommendation task. The main contribution of this paper is threefold:

- We applied the context tree structure to session-based recommendation tasks.
- Apart from the standard static evaluation protocol, we use an adaptive configuration where new items arrive continuously with limited observations. Extensive experiments and analyses are conducted on both configurations to benchmark CT with state-of-the-art methods.
- We show that CT outperforms other strong baseline methods in majority of the tested configurations and datasets. We analyze reasons for this and demonstrate that CT is particularly strong at recommending for relatively long sessions and adapting to new items with limited observations.

## 2 RELATED WORK

Among parametric methods for recommendation, the most well-known class of recommender system is based on matrix factorization (MF) [27]. Several attempts [18, 34] have been made to incorporate temporal components into the collaborative filtering setting to model users' preferences over time. Online versions of MF [11] are also widely studied for optimizing factorization-based methods efficiently. Session-based Matrix Factorization (SMF) is proposed very recently by [21] for the specific task of session-based recommendation. They also show that SMF constantly outperforms other factorization-based methods [10, 17] by large margin.

Recurrent neural networks (RNNs) [22] are recently applied successfully to session-based recommendation tasks [12, 13, 23]. The best-performing methods so far are based on GRU4Rec [13], and it is shown that they outperform other baselines by large margins in sequential recommendation tasks. Further improvements have been proposed by designing specific ranking loss and additional samples [12], or augmenting training data[29]. More details of methods based on GRU4Rec will be introduced in Section 4.1. A recent paper [3] proposes a Bayesian version of GRU4Rec with satisfying performance. They represent the network recurrent units as stochastic latent variables, and the corresponding posteriors are inferred using approximate variational inference for recommendation. It is demonstrated recently [33] that RNNs are suitable for collaborative filtering on dense datasets where state transition function is learned instead of the whole latent space. However, both MF and RNN break down in fully adaptive environments since matrix decomposition operation for MF and optimizing a large RNN are costly when data is large, which prevents them to be frequently updated.

Reinforcement learning methods are also proposed in several literatures for sequential recommendation task. [26] viewed the problem of generating recommendations as a Markov Decision Processes (MDPs), and they considered a finite mixture of Markov models with fixed weights. [24] proposed to factorize the transitions in Markov chains with low rank representation for basket recommendations. Their case only considers a first-order Markov chain, however, a higher order should be more interesting and realistic. [2] applied Markov models using skipping and weighting techniques for modeling long-distance relationships within a sequence. A major drawback of these Markov models is that it is not clear how to choose the order of Markov chain. A multi-armed bandit model called LinUCB is proposed by [20] for news recommendation to learn the weights of the linear reward function, in which news articles are represented as feature vectors; click-through rates of articles are treated as the payoffs. [31] proposed a similar recommender for music recommendation with rating feedbacks, called Bayes-UCB, that optimizes the nonlinear reward function using Bayesian inference. However, the exploration phase of these methods makes them adapt slowly. As user preferences drift fast in many recommendation settings, it is not effective to explore all options before generating useful ones.

Nonparametric methods, such as K-Nearest Neighbor (KNN) [30], are efficient and effective to deal with cold start problem in recommendation. For example, the item-to-item approach [13, 25] has been proven to be effective and is widely employed in industry. However, the representation power of nonparametric methods are limited due to simple prediction heuristics. Recently, [16] proposed a session-based nearest neighbor approach for session-based recommendation based on co-occurrence statistics from sessions of activities, and they show that the model is efficient to use with precomputed in-memory statistics and it achieves comparable performance against the RNN proposed by [13]. In this paper, we focus on a class of recommender systems based on a structure, called context tree [32], which was originally used to estimate variable-order Markov models (VMMs) for lossless data compression. [5, 19] applied this structure to various discrete sequence prediction tasks. Recently it was applied to news recommendation by [6, 7]. As one of the nonparametric methods, it can be updated and used efficient in an online fashion. An important property of online algorithms is the no-regret property, meaning that the model learned online is eventually as good as the best model that could be learned offline. According to [32], a no-regret property is achieved by context trees for the data compression problem. [5] showed through simulation that CT achieves a no-regret property when the environment is stationary. As we focus session-base recommendation tasks where data sparsity is often high and new items arriving continuously and limited observations, the no-regret property is hard to achieve while the model adaptability to new trends is a bigger issue for better performance.

## 3 CONTEXT TREE (CT) RECOMMENDER

Due to the sequential item consumption process, user preferences can be summarized by the last several items visited. When modeling the process as a fixed-order Markov process [26], it is difficult to select the depth of the Markov chain. A variable-order Markov model (VMM), like the context tree (CT), alleviates this problem by using a context-dependent depth.

In CT, a sequence $\mathbf{s} = \langle n_1, \ldots, n_l \rangle$ is an ordered list of items $n_i \in N$ consumed by a user. The sequence of items until time $t$ is $\mathbf{s}_t$ and the set of all possible sequences is $\mathcal{S}$. A context $S = \{\mathbf{s} \in \mathcal{S} : \xi \prec \mathbf{s}\}$ is the set of all possible sequences in $\mathcal{S}$ ending with the suffix $\xi$. $\xi$ is the suffix ($\prec$) of $\mathbf{s}$ if last elements of $\mathbf{s}$ match $\xi$. For example, $\xi = \langle n_3, n_1 \rangle$ is a suffix of sequence $\mathbf{s} = \langle n_2, n_3, n_1 \rangle$. A context tree $\mathcal{T} = (\mathcal{V}, \mathcal{E})$ with nodes $\mathcal{V}$ and edges $\mathcal{E}$ is a partition tree over all contexts of $\mathcal{S}$. Each node $i \in \mathcal{V}$ in the context tree corresponds to a context $S_i$. If node $i$ is the ancestor of node $j$ then $S_j \subset S_i$. Figure 1 illustrates a simple CT with some sequences over an item set $\langle n_1, n_2, n_3 \rangle$. Each node in the CT corresponds to a context. For instance, the node $\langle n_1 \rangle$ represents the context of all sequences that end with item $n_1$, and the node $\langle n_3, n_1 \rangle$ represents the context of all sequences that end with suffix $\langle n_3, n_1 \rangle$. Such suffix organization is used to capture the last couple of activities (context) of a user for session-based recommendation task.

**Figure 1: An example context tree. For the sequence $\mathbf{s} = \langle n_2, n_3, n_1 \rangle$, nodes in red-dashed are activated.**

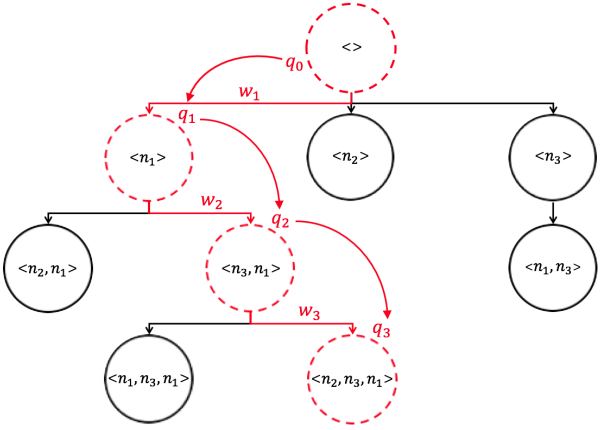

### 3.1 Local Experts for Context

For each context $S_i$, an expert $\mu_i$ is associated for computing the estimated probability $\mathbb{P}(n_{t+1}|\mathbf{s}_t)$ of the next item $n_{t+1}$ under this context. The standard way for estimating the probability $\mathbb{P}(n_{t+1}|\mathbf{s}_t)$, as proposed by [5], is to use a Dirichlet-multinomial prior for each expert $\mu_i$. The probability of viewing an item $x$ depends on the number of times $\alpha_{xt}$ this item has been consumed until time $t$ when the expert is active. The corresponding marginal probability is:

$$\mathbb{P}_i(n_{t+1} = x|\mathbf{s}_t) = \frac{\alpha_{xt} + \alpha_0}{\sum_{j \in \mathcal{N}} \alpha_{jt} + \alpha_0} \quad (1)$$

where $\alpha_0$ is the initial count of the Dirichlet prior.

### 3.2 Combining Experts into a Prediction

A user's browsing history $\mathbf{s}_t$ is matched to the context tree and identifies a path of matching nodes (see Figure 1), and all the experts associated with these nodes are *activated*. The set of *active* experts $\mathcal{A}(\mathbf{s}_t)$ for sequence $\mathbf{s}_t$ is the set of experts $\mu_i$ associated to contexts $S_i = \{\mathbf{s} : \xi_i < \mathbf{s}_t\}$ such that $\xi_i$ are suffix of $\mathbf{s}_t$. The set of activated experts $\mathcal{A}(\mathbf{s}_t)$ is responsible for the recommendation for $\mathbf{s}_t$.

The combined prediction of the first $i$ experts is defined as $q_i$ and it is computed using the recursion in Eq.2. The recursive construction estimates, for each context at a certain depth $i$, whether it makes better a prediction than the combined prediction $q_{i-1}$ from the active expert of depth $i-1$.

$$q_i = w_i \mathbb{P}_i(n_{t+1} = x|\mathbf{s}_t) + (1 - w_i) q_{i-1} \quad (2)$$

The weight $w_i$ is updated in closed form in Eq.3 by taking into account the success of a recommendation. When a user consumes an item $x$, the weights of all active experts are updated according to the probability $q_i(x)$ of predicting $x$ incrementally on the path via Bayes' theorem. More details and proofs of this design can be found in [5].

$$w'_i = \frac{w_i \mathbb{P}_i(n_{t+1} = x|\mathbf{s}_t)}{q_i(x)} \quad (3)$$

### 3.3 CT Recommendation Algorithm

The CT recommender is designed to be used in adaptive sequential recommendation environments. As new events in sequences (browser sessions) arrive continuously, the recommendation process consists of the following steps:

(1) match the history of the new event with the current tree to identify the active path of contexts and experts;
(2) expand node when new item or branch is encountered;
(3) Compute from the root to leaf along the active path;
   (a) update local experts by Eq.1;
   (b) update weights by Eq.3;
   (c) generate recommendation by Eq.2;

As more observations arrive, more contexts are added and updated, thus building a deeper and more complete tree. Moreover, it is a space efficient structure to keep track of the history in a variable-order Markov chain so that the data structure is built incrementally for sequences that actually occur. Therefore, it is a nonparametric method that can be used efficiently and robustly.

Two properties of the CT need to be noted. First, the model parameter learning process and recommendations generated can be computed efficiently in an online recursive manner such that the model adapts continuously and efficiently to adaptive environments. Second, adaptability can be achieved by the CT structure itself as knowledge is organized and activated by context. New items and paths can be immediately identified through context matching for fast adaptation, whereas old items can still be accessed in their old contexts. This allows the model to make predictions using more complex contexts as more data is acquired. Therefore, both old and new knowledge can be elegantly combined.

### 3.4 Complexity Analysis

For trees of depth $D$, the time complexity of model learning and prediction for a new observation are both $O(D)$. For an input sequence of length $T$, the updating and recommending complexity are $O(M^2)$, where $M = min(D, T)$. Therefore, both learning and prediction are efficient with the CT. Space complexity in the worst case is exponential in the depth of the tree. However, as we do not generate branches unless the sequence occurs in the input, we achieve a much lower bound controlled by the total size of the input. So the space complexity is $O(N)$, where $N$ is the total number of observations.

## 4 STATE-OF-THE-ART METHODS

This section reviews two categories of best-performing methods so far for session-based recommendation tasks.

Table 1: Statistics for all datasets. Training and test data are used in static configuration; full data is used in adaptive configuration.

|  | Full Data | | | | Training Data | | | Test Data | |
| --- | --- | --- | --- | --- | --- | --- | --- | --- | --- |
|  | Items | Events | Sessions | Avg. Session Len. | Items | Events | Sessions | Events | Sessions |
| *Course 1* | 1,116 | 125,077 | 14,876 | 8.4 | 1,068 | 109,988 | 13,788 | 12,941 | 726 |
| *Course 2* | 1,643 | 359,414 | 20,602 | 17.4 | 1,592 | 329,938 | 19,357 | 28,385 | 1,129 |
| *Course 3* | 2,403 | 772,412 | 25,191 | 30.7 | 2,368 | 665,310 | 23,894 | 106,753 | 1,287 |
| *News 1* | 4,040 | 57,182 | 18,099 | 3.2 | 3,919 | 53,363 | 16,963 | 2,389 | 709 |
| *News 2* | 6,358 | 955,961 | 205,988 | 4.6 | 2,285 | 876,077 | 202,656 | 58,451 | 3,054 |
| *RSC15* | 37,483 | 31,708,461 | 7,981,581 | 4.0 | 37,483 | 31,637,239 | 7,966,257 | 71,222 | 15,324 |
| *XING* | 59,297 | 546,862 | 89,591 | 6.1 | 58,035 | 488,576 | 78,276 | 58,286 | 11,315 |

## 4.1 Recurrent Neural Networks

Recurrent Neural Networks (RNNs) have been applied extensively to sequence learning tasks. Figure 2 illustrates a general structure of using RNNs for session-based recommendation tasks. The major component of RNNs is the hidden state captured in the recurrent layer. Special recurrent units like Long Short-term Memory (LSTM) [14] and Gated Recurrent Units (GRU) [4] are widely used for modeling long-term dependencies within a sequence with the help of specialized gate functions, and they often help to boost performance significantly. The input of the network is the item that a user currently views in a session while the output is the next item she views in the session. Furthermore, a embedding layer can be used to learn a compact representation of each item before feeding to recurrent layers. A feed-forward layer will be used to map the output of recurrent layer to the dimension of output labels. Both input and output layers use one-hot vector encodings, that is, the vector length equals to the number of items and only the coordinate corresponding to the active item is one, the others are zeros. Due to the limitation of the one-hot vector representation, when a new item arrives, one more dimension need to be added. Therefore, the whole parametrized network, from the input layer to the output layer, needs to be adjusted to the new item, which is very costly and data-intensive to train.

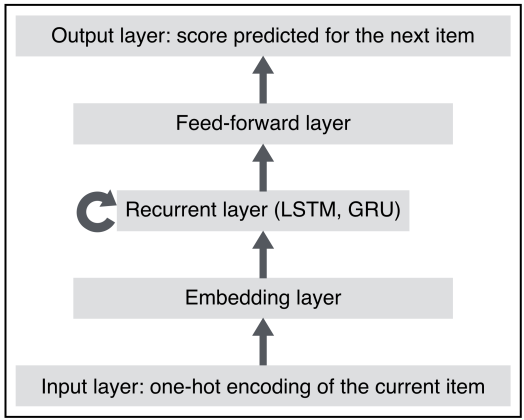

Figure 2: General structure of RNNs for session-based recommendation. The figure is adapted from [13]

A typical loss function in this case is the cross-entropy loss, which measures the distance between the predicted distribution $p$ over labels and the distribution $q$ of true labels (one-hot vector) by:

$$H(p,q) = -\sum_{j=1}^{N} q_j \log p_j = -\log p_i \qquad (4)$$

where $N$ is the number of labels and $p_i$ is the predicted value on the desired item $i$. The state-of-the-art RNN for session-based recommender is referred to GRU4Rec [13]. Their model follows the general structure in Figure 2, with GRU as recurrent cells. They additionally used the relevance-based ranking of items to define two different ranking losses by:

$$Top1 : L_s = -\frac{1}{N_S} \sum_{j=1}^{N_S} \sigma(p_j - p_i) + \sigma(p_j^2) \qquad (5)$$

$$BPR : L_s = -\frac{1}{N_S} \sum_{j=1}^{N_S} \log(\sigma(p_j - p_i)) \qquad (6)$$

where $N_S$ is the size of negative samples used to compute the ranking loss. Moreover, they propose a session-parallel minibatch technique for optimizing variable-length sequences. A more recent version, GRU4Rec+ [12], explored different sophisticated ranking losses to have the target score compared with the most relevant negative samples only. They also used an improved sampling strategy with additional negative samples.

## 4.2 K-Nearest Neighbor Approach

The K-Nearest Neighbor (KNN) approach finds the k most similar items in a training set to a particular testing point based on a given distance metric. It is a commonly used machine learning method for pattern recognition[35] and information retrieval [15]. KNN is also broadly applied to recommendation task due to its simplicity. For session-based recommendation task, the simplest KNN method, called Item-KNN, is benchmarked by [13]. Item-KNN generates recommendations that are the most similar to the single last item in a given session, and similarity is defined by item co-occurrence statistics across sessions. Recently, an improved version of KNN, called session-based KNN (SKNN), is proposed by [16] for session-based recommendation by considering similar sessions. SKNN takes items a user viewed in the current session $s$ and determines the $k$ most similar past sessions $N_s$ in the training data. Then, the score of a recommendable item $i$ is computed as:

$$score_{SKNN}(i, s) = \sum_{n \subset N_s} sim(s, n) \times 1_n(i) \qquad (7)$$

where $1_n(i) = 1$ if session $n$ contains $i$ and 0 otherwise. The similarity function $sim(s, n)$ computes the similarity between two sessions *s* and *n* by cosine similarity with both sessions encoded as binary vectors of the item space. [16] compared SKNN with GRU4Rec. They show that SKNN has comparable performance against GRU4Rec, and the model is efficient to use with precomputed in-memory statistics. SKNN does not consider the order of item in a session. A recent work by [21] proposed a sequence-aware SKNN (S-SKNN) with more weights on items that appear later in the session.

## 5 EXPERIMENTS

### 5.1 Dataset Description

**MOOC Datasets:** The first set of three datasets come from discussion forum view data of three MOOCs offered by our university on Coursera[1], referred to *Course 1*, *Course 2* and *Course 3*. [2]. Items to be recommended are discussion forum threads that contain several posts and comments within the topic. These three courses last for around ten to twelve weeks and interactions from the last week are assigned to test set for static evaluation. At the beginning of a MOOC, the discussion forum is empty and all threads are developed along the span of the course. Furthermore, student interests are influenced by the progress of a course and items themselves are evolving over time because forum threads can be created or edited very frequently by either students or instructors, therefore, recommendations need to be adaptive to these drifting preferences and evolving items, which makes the MOOCs dataset a perfect choice for the later adaptive configuration.

**News Datasets:** The next two datasets, referred as *News 1* and *News 2* [3], are in the domain of news recommendation. They are collected from "recommendations-free" clicks log of two local news websites for the span of three weeks and three days respectively, where *News 2* is a much busier site. The last-day and last-hour interactions of the two news datasets are assigned to test sets for static evaluation.

**RecSys Challenge Datasets:** The last two datasets are used by state-of-the-art RNNs [12, 13, 23, 29] and KNNs [16, 21] for session-based recommendation. RSC15 is from RecSys Challenge 2015[4], which contains click-streams of an e-commerce site during the span of 6 months. The other one, called XING, is from the Recsys Challenge 2016[5] and contains interactions on a job postings website over a 80-day period. Activities from sessions in the last day of the two datasets are assigned to test sets

In all of our datasets, only sessions with length longer than two are retained. For MOOC and news datasets, duplicate items in a session are kept, while for the two RecSys Challenge datasets, as in [12, 13, 16, 23, 29], duplicate items are removed and infrequent items with less than five views are also discarded. Statistics of the seven datasets are summarized in Table 1.

### 5.2 Evaluation Metrics

Our session-based recommendation task is to predict the immediate next item given the previous interactions in a session. The recommendation model generates a ranked list of *k* items as prediction for each testing interaction event, and the evaluation metrics that we are going to use across later experiments are as follow:

- **HR@k**: The averaged hit rates (recall) of having the desired items amongst the top-k recommended item lists.
- **MRR@k**: The averaged mean reciprocal ranks of the desired items in top-k recommended item lists.

### 5.3 Static Evaluation Configuration

In this static setting, models are trained on a training set and evaluated on another static test set that corresponds to events that arrive after a certain point. This static configuration is adopted by all previous researches [12, 13, 16, 21, 23] for session-based recommendation task. The major drawback of such static configuration is that the test set only contains items that appeared during the training phase. Furthermore, strict preprocessing steps are often conducted. Those assumptions do not apply in many real-life applications, therefore, such static configuration is mainly used to study the sequence learning capability of different methods.

In a static evaluation configuration, CT is benchmarked against two state-of-the-art categories of session-based recommendation algorithms using RNNs or KNNs described in Section 4.1 and 4.2. Another compared algorithm, called Session-based Matrix Factorization (SMF) [21], is proposed very recently for the specific task of session-based recommendation. It is shown that it is constantly among the best-performing methods and it outperforms other factorization-based methods [10, 17] by large margins. In SMF, the traditional user latent vector is replaced by a session specific latent vector and model is optimized using gradient descent using the complicated ranking loss proposed by [12]. For methods based on RNNs or SMF, it is extremely time-consuming to optimize all hyperparameters for all datasets. Therefore, we fix hyperparameters as summarized in Table 2, and only select dropout and learning rates using grid search on corresponding validation sets. For two versions of GRU4Rec, single recurrent layer is used as in previous works. For CT, the longest context length is set to 50, and no other hyperparameters are required. For SKNN, 500 nearest neighbors from the 1000 most recent candidate sessions are used while 100 nearest neighbors from the 500 most recent candidate sessions are considered for S-SKNN.

*5.3.1 Static Evaluation Results.* The results for the three MOOC datasets are presented in Table 3. We could see that CT performs the best as it is only outperformed by GRU4Rec+ on one metric. Two versions of GRU4Rec and SMF form the second tier with GRU4Rec+ performs better. For KNN-based methods, simple ItemKNN performs slightly better than two other session-based KNNs on these three datasets. However, this is in general not true when considering other news and RecSys Challenge datasets in Table 4 where two session-based KNNs performs much better than ItemKNN. Furthermore, ItemKNN fails on News 2 where users tend to view an item repeatedly. Results in Table 4 for news and RecSys Challenge datasets show that CT is still the best-performing methods as it wins eight out of sixteen evaluation metrics while S-SKNN and GRU4Rec go for the second-best with each wining four and three evaluation metrics. Some conclusions can be drawn from the static evaluations in both Table 3 and Table 4 :

- CT performs the best across different datasets and evaluation metrics. It is a strong signal that CT learns sequential patterns

---
[1] coursera.org
[2] More details of MOOCs datasets will be provided if the paper was accepted
[3] More details news datasets will be provided if the paper was accepted
[4] http://2015.recsyschallenge.com/
[5] http://2016.recsyschallenge.com/

**Table 2: Hyperparameters used for different algorithms on all datasets except RSC15. Dropout rate and learning rate are selected by grid search on validation sets for seven datasets. For manageable computational efficiency on the largest RSC15 dataset, the recurrent layer sizes of GRU4Rec and GRU4Rec+ are set to be 100 instead of 1000.**

|         | Loss    | Hidden Size | Activation Function | Momentum | Regularization | Additional Samples |
|---------|---------|-------------|---------------------|----------|----------------|--------------------|
| GRU4Rec | Top1    | 1000        | tanh                | 0.1      | -              | -                  |
| GRU4Rec+| BPR-Max | 1000        | tanh                | 0.1      | -              | 2048               |
| SMF     | Top1-Max| 100         | linear              | 0.2      | 0.005          | 2048               |

**Table 3: Results for static evaluation on three MOOC datasets. Models in the table are ordered by the averaged performance across different metrics and the best method in each column is highlighted.**

|         | *Course 1* | | | | *Course 2* | | | | *Course 3* | | | |
|---------|-------|--------|-------|--------|-------|--------|-------|--------|-------|--------|-------|--------|
|         | HR@5  | MRR@5  | HR@20 | MRR@20 | HR@5  | MRR@5  | HR@20 | MRR@20 | HR@5  | MRR@5  | HR@20 | MRR@20 |
| CT      | **0.369** | **0.233** | **0.578** | **0.254** | **0.343** | **0.219** | 0.549 | **0.240** | **0.355** | **0.224** | **0.591** | **0.248** |
| GRU4Rec+| 0.358 | 0.217 | 0.573 | 0.242 | 0.334 | 0.195 | **0.574** | 0.221 | 0.328 | 0.194 | 0.586 | 0.222 |
| SMF     | 0.345 | 0.207 | 0.570 | 0.230 | 0.314 | 0.184 | 0.543 | 0.207 | 0.329 | 0.198 | 0.579 | 0.225 |
| GRU4Rec | 0.334 | 0.198 | 0.554 | 0.220 | 0.274 | 0.165 | 0.491 | 0.185 | 0.294 | 0.178 | 0.512 | 0.200 |
| ItemKNN | 0.269 | 0.158 | 0.473 | 0.178 | 0.216 | 0.123 | 0.405 | 0.142 | 0.227 | 0.135 | 0.446 | 0.156 |
| SKNN    | 0.252 | 0.140 | 0.441 | 0.159 | 0.243 | 0.135 | 0.379 | 0.150 | 0.181 | 0.101 | 0.273 | 0.118 |
| S-SKNN  | 0.273 | 0.160 | 0.447 | 0.178 | 0.229 | 0.133 | 0.325 | 0.144 | 0.149 | 0.093 | 0.149 | 0.100 |

**Table 4: Results for static evaluation on news and RecSys Challenge datasets ([†] Slightly worse than [13]; [‡] Better than [23] [★]; Taken from [12, 21]; [§] Consistent with [16, 21]). Models in the table are ordered by the averaged performance across different metrics and the best method in each column is highlighted.**

|         | *News 1* | | | | *News 2* | | | | *RSC15* | | | | *XING* | | | |
|---------|-------|--------|-------|--------|-------|--------|-------|--------|-------|--------|-------|--------|-------|--------|-------|--------|
|         | HR@5  | MRR@5  | HR@20 | MRR@20 | HR@5  | MRR@5  | HR@20 | MRR@20 | HR@5  | MRR@5  | HR@20 | MRR@20 | HR@5  | MRR@5  | HR@20 | MRR@20 |
| CT      | **0.465** | **0.314** | 0.595 | **0.329** | 0.847 | **0.795** | 0.898 | **0.800** | 0.457 | **0.287** | 0.669 | 0.306 | 0.158 | **0.110** | 0.199 | **0.115** |
| GRU4Rec+| 0.457 | 0.284 | **0.646** | 0.305 | 0.859 | 0.766 | **0.948** | 0.776 | 0.431 | 0.277 | **0.719**[★] | **0.312**[★] | 0.156 | 0.094 | 0.245 | 0.103 |
| S-SKNN  | 0.457 | 0.299 | 0.626 | 0.318 | **0.885** | 0.790 | 0.931 | 0.794 | 0.409 | 0.238 | 0.633[§] | 0.262[§] | **0.184** | 0.109 | **0.284** | 0.109 |
| SKNN    | 0.438 | 0.293 | 0.633 | 0.314 | 0.857 | 0.754 | 0.929 | 0.762 | 0.390 | 0.245 | 0.641[§] | 0.250[§] | 0.165 | 0.102 | 0.276 | 0.113 |
| SMF     | 0.435 | 0.269 | 0.565 | 0.284 | 0.829 | 0.718 | 0.925 | 0.728 | **0.476**[★] | 0.284[★] | 0.712[★] | 0.309[★] | 0.146 | 0.084 | 0.228 | 0.092 |
| GRU4Rec | 0.441 | 0.280 | 0.570 | 0.293 | 0.833 | 0.756 | 0.900 | 0.763 | 0.347 | 0.237 | 0.615[†] | 0.269[†] | 0.148[‡] | 0.090[‡] | 0.230 | 0.098 |
| ItemKNN | 0.277 | 0.190 | 0.435 | 0.211 | 0.073 | 0.045 | 0.157 | 0.055 | 0.271 | 0.177 | 0.507 | 0.205 | 0.070 | 0.041 | 0.126 | 0.045 |

- as well as a fully optimized RNN in static environments. Moreover, it in general performs better on the three MOOC datasets where sessions are longer.
- Two versions for GRU4Rec give very good performance as well, with GRU4Rec+ consistently better than GRU4Rec. Moreover, SMF has very comparable performance against the basic GRU4Rec in majority of the evaluation metrics.
- Session-based KNNs (SKNN and S-SKNN) perform very well for news and RecSys Challenge dataset, but not good enough for three MOOC datasets.

*5.3.2 Computational Efficiency.* Table 5 shows the training time and response time per recommendation of different methods used in static evaluation on four largest datasets. All methods are computed using a machine with 2x12cores@2.5 GHz Intel Core i5-3210M processor. Two versions of GRU4Rec and SMF are trained on two NVIDIA TITAN X GPUs with 12GB memory on the same machine. We can note that different nonparametric methods are extremely fast to train, with SKNN to be the most efficient. Compared with SKNN, CT take slightly more time to train and predict. Compared to the slighted increased computation cost, we contend that the performance boost still support the usability of CT. Compared to nonparametric methods, both versions of GRU4Rec and SMF take extremely long time to train. Therefore, we contend that the performance improvement of using GRU4Rec+ over GRU4Rec or CT in some cases does not compensate for the dramatically increased computation cost.

### 5.4 Adaptive Evaluation Configuration

In this adaptive evaluation configuration, we replay interaction events that arrive one-by-one by time order and the recommender is updated continuously. Recommendations are generated using the continuously updated model and the goal is to be adaptive to new items and patterns with limited observations at prediction. This is the practically most important setting in real-life applications and a favorable recommender needs to deal with such context effectively. Different versions of neighborhood-based methods and CT can be easily implemented for this adaptive configuration thanks to efficient computation, while methods based on matrix factorization or RNNs need to be modified correspondingly for the task. We included a online version of matrix factorization (Online-MF) [11] with implicit user feedback. The number of latent factor is set to 16 and the model is updated using the efficient element-wise ALS proposed by [11]. GRU4Rec here is updated after a session (sequence) of events arrives, as it is not computationally feasible to update an RNN upon the arrival of each single event. As it is by nature not efficient to use GRU4Rec for adaptive configuration, we do not try to fully optimize GRU4Rec but rather highlight the performance gap between using RNNs and nonparametric methods. Furthermore, some other baselines based on simple heuristics are also included: "Random" recommends 5 random items among last 100 viewed ones; "Fresh" recommends 5 most freshest item; "Popular" recommends 5 top-popular items among last 100 viewed ones;

**Table 5: Computational complexity for four largest datasets in terms of total training time and average prediction time per recommendation. On the largest RSC15 dataset, the reported computation times of GRU4Rec (+) correspond to hidden layer size of 100, rather than 1000 as in other datasets.**

|  | *Course 3* | | News2 | | RSC15 | | XING | |
|---|---|---|---|---|---|---|---|---|
|  | Train. (min) | Pred. (ms) | Train. (min) | Pred. (ms) | Train. (min) | Pred. (ms) | Train. (min) | Pred. (ms) |
| SKNN | 0.03 | 7.2 | 0.06 | 27.2 | 1.1 | 35.1 | 0.04 | 10.9 |
| CT | 1.9 | 22.5 | 1.1 | 21.7 | 34 | 57.0 | 0.9 | 74.1 |
| SMF | 173 | 12.8 | 77 | 7.2 | 409 | 24.3 | 89 | 19.8 |
| GRU4Rec | 267 | 26.7 | 283 | 25.6 | 299 | 21.1 | 176 | 35.0 |
| GRU4Rec+ | 755 | 27.1 | 924 | 26.0 | 1890 | 25.1 | 375 | 36.4 |

**Table 6: Results for adaptive evaluation with *tail performance* metric in parentheses. Models in the table are ordered by the averaged performance across different metrics and the best method in each column is highlighted.**

|  | **Course 1** | | **Course 2** | | **Course 3** | | **News 1** | | **News 2** | |
|---|---|---|---|---|---|---|---|---|---|---|
|  | HR@5 | MRR@5 | HR@5 | MRR@5 | HR@5 | MRR@5 | HR@5 | MRR@5 | HR@5 | MRR@5 |
| CT | **0.381 (0.295)** | **0.260** | **0.354 (0.179)** | **0.228** | **0.321 (0.238)** | **0.207** | 0.522 (0.143) | 0.355 | **0.677 (0.163)** | **0.562** |
| SKNN | 0.327 (0.270) | 0.229 | 0.332 (0.142) | 0.192 | 0.287 (0.204) | 0.167 | **0.590 (0.210)** | **0.372** | 0.664 (0.163) | 0.545 |
| GRU4Rec | 0.296 (0.210) | 0.194 | 0.304 (0.101) | 0.156 | 0.262 (0.180) | 0.142 | 0.370 (0.067) | 0.211 | 0.604 (0.110) | 0.417 |
| Popular | 0.197 (0.142) | 0.106 | 0.238 (0.096) | 0.130 | 0.197 (0.135) | 0.105 | 0.278 (0.006) | 0.152 | 0.285 (0.005) | 0.182 |
| Online-MF | 0.184 (0.133) | 0.114 | 0.155 (0.091) | 0.109 | 0.117 (0.092) | 0.084 | 0.281 (0.059) | 0.166 | 0.339 (0.070) | 0.202 |
| Fresh | 0.126 (0.120) | 0.051 | 0.065 (0.062) | 0.025 | 0.068 (0.063) | 0.028 | 0.201 (0.003) | 0.080 | 0.035 (0.014) | 0.017 |
| Random | 0.022 (0.016) | 0.011 | 0.011 (0.008) | 0.006 | 0.010 (0.007) | 0.005 | 0.125 (0.008) | 0.061 | 0.160 (0.008) | 0.089 |

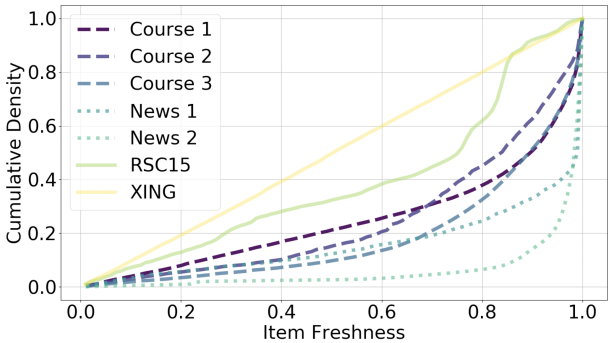

**Figure 3: ECDFs of distribution on item freshness for all datasets**

*5.4.1 Item View Patterns.* Figure 3 presents the empirical cumulative density functions (ECDFs) of the distribution of *item freshness* for seven datasets. When an item is viewed, the *item freshness* of this event is the relative creation order of the item. For example, if an item is the $m$-th created one among the current pool of $n$ items, the *item freshness* for this event is $\frac{m}{n}$. We can see that the most adaptive pattern appears in two news datasets where new items are viewed much more frequently than old ones. A similar pattern is observed for three MOOC datasets. This dynamic pattern mainly comes from the fact that new items are more interesting for users in these two domains. The other two datasets (RSC15 and XING) require less adaptation. Therefore, the first five datasets (MOOCs and News) are used for adaptive evaluation configuration. Although the two news datasets look the most dynamic, they only contain data from a limited period of time where fresh items are not necessarily created very recently. In contrast, each MOOC dataset individually contains an entire life span of a small community with not item at the very beginning. Such a cold-start configuration is rather crucial for the application of session-based recommendation techniques to small retailers or to domains with evolving items, such as discussion forums or social media platforms. Therefore, we contend that the performance on three MOOC dataset to be the most important and interesting for adaptive configuration.

*5.4.2 Adaptive Evaluation Results.* The results in Table 6 show the performance of various methods on five datasets for adaptive evaluation, with results in parentheses consider a *tail performance* metric, where the most popular items are excluded from recommendations. We can see that the CT recommender outperforms all other sequential methods on three MOOC datasets and News 2. SKNN performs the best on News 1 dataset and it performs the second best on other datasets. For two news datasets, the average session length is very short according to Table 1, which means that users do not visit a large number of items in a single session. In such cases, the relatively simple SKNN outperforms CT which mines sequential patterns. However, we could see a clear win of CT over SKNN when session length is long, as in the three MOOC datasets. The performance of GRU4Rec is slightly worse than SKNN but consistently better than Online-MF, and GRU4Rec in general performs better when session length it longer. The Online-MF method, together with the Popular recommender, forms the next tier. Online-MF performs better than the Popular recommender on two news datasets but not on the three MOOCs datasets. Furthermore, the performance gap of these two methods against the top three models is very big. These result shows that nonparametric methods, such as CT and SKNN, that can be updated efficiently and effectively are suitable for adaptive evaluation configuration. CT in general performs better than SKNN, especially for three MOOC datasets with aforementioned complete item life cycle and long session length. Moreover, both RNNs and MF are not adaptive enough to patterns in new items with limited observations, with GRU4Rec perform better than Online-MF.

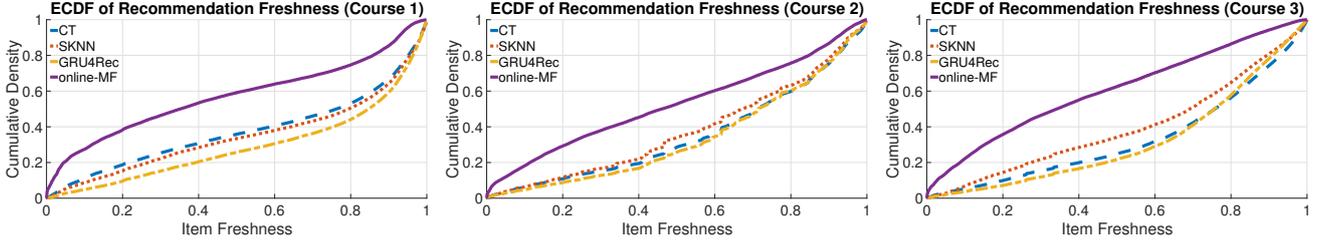

Figure 4: ECDFs of recommendation freshness of CT, SKNN, GRU4Rec, and Online-MF for MOOCs datasets

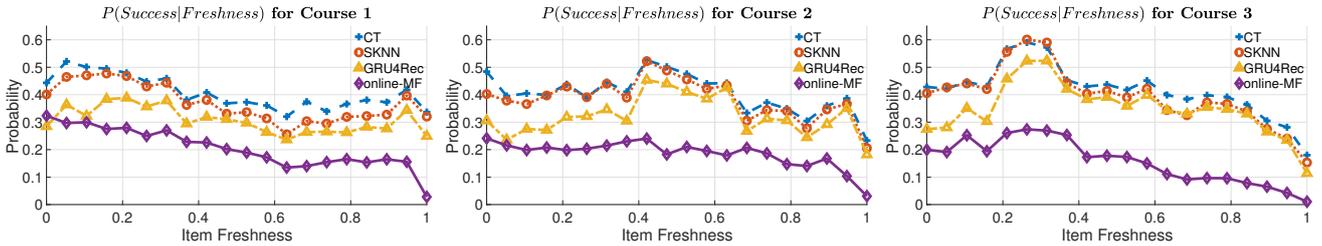

Figure 5: Conditional success rate of CT, SKNN, GRU4Rec, and Online-MF for MOOCs datasets

#### 5.4.3 Adaptation Comparison.
After seeing the distinguished performance of the CT recommender in adaptive evaluation configuration, we move to an insight analysis of the result. To be specific, we compare CT, SKNN, GRU4Rec, and online-MF in terms of their adaptation capabilities to new items on three MOOCs datasets. Figure 4 is a quantitative comparison that illustrates the empirical cumulative density functions (ECDFs) of items recommended by different methods against item freshness. We can see that the ECDFs of CT, SKNN and GRU4Rec increase sharply as item freshness increases, which means that they recommend much more new items than Online-MF. Figure 5 shows qualitative results of the conditional success rate $P(Success|Freshness)$ against freshness. We can see that Online-MF fails badly when freshness is high. In contrast, CT, SKNN and GRU4Rec maintain relatively stable performance across all levels of freshness with CT and SKNN consistently better than GRU4Rec across all levels of item freshness. We believe that the data sparsity issue in adaptive environments hinders the performance of RNNs due to insufficient model training on new items. CT performs similarly compared to SKNN when item freshness is low, yet it is consistently better than SKNN on fresh items. This result, congregated with findings from previous adaptive evaluation in Table 6, reinforced our conclusion that CT is versatile for both old and new items and it is particularly strong at recommending fresh items with limited observations.

We conclude by summarizing the experimental results from both static and adaptive configurations:

- In static configuration, CT learns sequential patterns as well as fully-optimized RNNs and it achieves top performance.
- Nonparametric methods are extremely fast to train and update. Therefore, well-devised methods, like CT, are suitable for adaptive real-life systems.
- Adaptive evaluation confirms that CT is versatile for both old and new items and it is particularly strong at recommending for relatively long sessions and adapting to new items with limited observations.

## 6 CONCLUSION AND FUTURE WORK
In this paper, we benchmark the CT recommender with the state-of-the-art methods based on MFs, RNNs and KNNs for both static and adaptive configurations. Through extensive experimental analyses, we show that the CT structure is well suited to learn sequential patterns and it is especially strong at adapting to new items. Furthermore, we stress the efficiency of using nonparametric methods to deal with adaptive domains in industry. As a future work, we will try to congregate the virtues of nonparametric methods and neural networks to achieve better performance and applicability.